# A prototype for a serious digital game to teach linguistic ontologies.

*Diana Medina, Grissa Maturana, Fernán Villa*
*dijmedinago@unal.edu.co, gvmatura@unal.edu.co, favillao@unal.edu.co*

**Working Paper**

**Abstract**

The objective of ontologies is to increase the compression of a given domain by eliminating interpretation problems. Among kinds of ontologies are linguistics ontologies which are ontologies used to simplify the interface between domain knowledge and linguistic components. Digital games have received increasing interest from educators in recent years for their potential to enhance the language learning and linguistic learning experience. Within the literature are games to teach ontologies of a specific domain, and games that use ontologies to facilitate the understanding of a given domain. Other educational games teach linguistics or vocabulary in contexts in which language is useful and meaningful. Although games help to understand difficult topics, the use of games that seek to meet the learning objectives of linguistics is not very popular and those focused on teaching linguistic ontologies are scarce. To solve the lack of the recreational resource for teaching linguistics in this document a prototype of a digital game called onto-ling is proposed. The goal is for the player to learn the relationship between concepts according to semantics, types of concepts and relationships through a game of levels.

**Keywords**

Linguistic, Ontology, Linguistics ontologies, Digital Game, Learning.

1. **Introduction**

One of the main goals of ontologies is to increase shared understanding in each domain. In this way, ontologies can function as a framework that unifies different viewpoints and improves communication (Aguado et al., 1998). Ontologies provide the basis for automated reasoning in a domain by declaring which entities exist and what types of relationships exist between those entities. It only requires understanding what terms mean, and what can be inferred from their use contexts, something that linguistics students should trained to do routinely (Farrar & Langendoen, 2003).

Linguistic ontologies can be seen both as a kind of lexical database and as a kind of ontology. Linguistic ontologies differ primarily from formal ontologies regarding their degree of formalization. Linguistic ontologies, in fact, do not reflect all the inherent aspects of formal ontologies (Magnini & Speranza, 2002), these are used to simplify the interface between domain knowledge and linguistic components, it offers a classification of the type of meanings that grammatical constructions presuppose (Aguado *et al.*, 1998).

As interactive, multimodal, immersive, and extremely popular environments, digital games have received increasing interest from educators in recent years for their potential to enhance the language learning experience, both inside and outside the classroom (Reinders H., 2017). Games can be successfully used to address different types of learning as well as a variety of subjects (Barreira *et al.*, 2012).

Within the literature, digital games have been set up to teach ontologies of a specific domain, such as The CIDOC CRM game (Guillem & Brusekerb, 2017), Mingoville (Hansbøl & Meyer, 2011) and others, and games that use ontologies as a way of representing information, including ITSEGO (Centola & Orciuoli, 2016), eMedOffice (Hannig *et al.*, 2012), and matchballs (Ziebarth *et al.*, 2012). Some educational games teach linguistics or vocabulary in contexts in which language is useful and meaningful, *eg.*, MOW (Barreira *et al.,* 2012). Although games help to understand different kinds of topics including difficult topics, the use of games that seek to meet the learning objectives of linguistics is not very popular in especial those focused on teaching linguistic ontologies are scarce.

A digital game called onto-ling is proposed, which seeks to reduce the lack of recreational resources as games for the teaching of linguistics, using linguistic ontologies. Onto-ling is a game based on Visuwords™, an interactive visual dictionary (https://visuwords.com/). The goal is for the player to learn the relationship between the concepts according to the semantics and the context, so he obtains the ability to differentiate the types of concepts and relationships through a set of levels.

In addition to the proposal, a base case with 5 basic levels is applied to a group of linguistic students is presented. We have compared the results from using Onto-ling along with traditional teaching methods. The results indicate that students who used the game had a superior learning progress than those who used only traditional methods.

The reminder of this paper is organized as follows: in Section 2 we present some theoretical framework related to ontologies, linguistic and linguistics ontologies. Current state-of-the-art ontologies games and linguistic games are presented in Section 3. Then, in Section 4 we propose a Onto-ling game, and we describe the rules and conditions of the game and the results of the game application. Finally, we present the conclusions and future work.

2. **Theoretical framework**

The term ontology originates from philosophy and it is the explanation of being; today it is used in computer science and knowledge engineering. The most common definition in literature has been coined by Struder *et al.* (1998), who defines ontology as "a formal explicit specification of a shared conceptualisation". In simpler terms, an ontology is a knowledge model that contains a group of concepts/terms that describe a specific domain (Trokanas et al., 2014).

Linguistic ontologies are large scale lexical resources that cover most words of a language, while at the same time also providing an ontological structure where the main emphasis is on the relations between concepts (Magnini & Speranza, 2002). The linguistic ontologies are used to simplify the interface between domain knowledge and linguistic components, offering a classification of the kind of meanings that grammatical constructions presuppose (Aguado et al., 1998).

More recently, the role of linguistic ontologies is also emerging in contexts where the problem of meaning negotiation is crucial. A relevant perspective in this direction is represented by linguistic ontologies with domain specific coverage, whose role has been recognized as one of the major topics in many application areas (Magnini & Speranza, 2002). Thus, that play a main role providing semantics for domain concepts and connecting conceptual representations with lexical representations (Aguado et al., 1998).

3. **Related work**

The CIDOC CRM game as proposed by Guillem & Brusekerb (2017), has a purpose of which is a learning mechanism to allow learners to approach CIDOC CRM, a formal ontology for medium and long-term integration of cultural heritage data to allow greater valorization and dissemination. The CIDOC CRM game consist of decks of cards and game boards that allow players to engage with the concepts of a formal ontology in relation to real data in an entertaining and informative way.

Hansbøl, & Meyer (2011), take as its point of departure a language project to serious games, and use Mingoville (http://www.mingoville.com) a virtual universe provides an online environment for students beginning to learn English in schools and at home. Hansbøl, & Meyer's (2011) proposal focuses on the shifting ontologies of Mingoville and teaching and learning situations in beginners' English. The arguments and descriptions provided in Mingoville is focus on shifting ontologies as it moves into, and out of, different teaching and learning situations of English for beginners.

Centola & Orciuoli (2016), propose the definition of a tool supporting the transition of children from kindergarten to primary school and, the development digital competences. The tool has been defined, by means of an ontology-driven approach as an Intelligent Tutoring System (ITS) integrated to a structured game based educational environment and provides benefits for both teachers and children. The definition of a novel ontology, namely ITSEGO, provides a model to build Game-based ITS for supporting the transition.

A computer-aided serious game (eMedOffice) developed and currently in use at the RWTH Aachen University Medical School is presented by Hannig *et al.* (2012). The game is part of the attempt to teach medical students the organizational and conceptual basics of the medical practice in a problem-based learning environment. Acquired domain knowledge was represented and formalized in an ontology to structure and maintain the knowledge base, ontology permits a flexible inclusion is provided of different types of rules like necessary distance between objects or required access from a certain direction.

A multi-agent-architecture for collaborative, serious and casual games is presented by Ziebarth *et al.* (2012). To be flexible concerning the learning domain an ontology-based approach has been used. The ontology may easily be exchanged to adapt the game to another domain. To test the architecture, an ontology on food safety and hazardous material regulations was created.

Barreira *et al.* (2012), present MOW (Matching Objects and Words), which is an Augmented Reality (AR) game developed in collaboration with elementary teachers that allows children to learn a variety of words in different languages, through a very first visual contact and oral comprehension, followed by an ongoing recognition and verbal domain of written words, much supported by memory as it is exercised through specific tasks within the game, providing to young children a full training and exploration of their inner abilities and learning capabilities. Also interesting is the fact that this game provides the teachers a didactic to support the teaching processes concerning the different linguistic topics.

According to analyzed games of current literature, the proposals of games whose main objective is to teach linguistic ontologies are scarce. There is the possibility of creating and providing an effective and fun learning methodology for linguistic ontologies. As a starting point it would be interesting to propose a game that allows the understanding of the linguistic components in relation to specific domains mainly to linguistic students.

4. **Game description and application**

We propose Ontol-ling as a digital game based on the well-known interactive visual dictionary Visuwords™. While visuword has educational and creative objectives, onto-ling has educational and evaluative objectives. Visuwords™ permits explore the lexicon. Whether the user is a native English speaker or a second language user—either as a student or teacher—he can use the graphs to associate words and expand on concepts. The user can move beyond synonyms and find other types of relational connections.

The goal of Onto-ling is for the player to learn the relationship between terms according to semantics, types of terms and relationships through a game by levels. After playing the player may be able to match words according to their meaning and identify the types of relational connections.

Onto-ling is suggested to be played in an individual way. The player receives in each level some definitions located in a network of nodes or 'synsets'. A synset is essentially a single concept that is represented by several terms or synonyms. Synonyms are words with different spellings that convey the same idea. Some synsets will also show a few examples of usage.

In each synset, the player must locate the term that corresponds to the respective definition considering that the kind of association fits the created expression. Once the player locates all the terms he will receive a score and he will be able to move on to the next level. The teachers can use the score as an evaluation score.

The game elements are represented by using the Visuwords™ representation. Each synset node is shown as a globe. Nouns are blue, verbs are green, adjectives; orange and adverbs; red. The synsets are joined by colored links that represent kind of association those synsets have to one another. The representation and description of game elements are included in the exhibit 1.

**EXHIBIT 1. The game elements. The authors.**

| Element | Description | Element | Description |
|---|---|---|---|
| 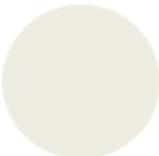 Definition | An explanation of the meaning of a term. | 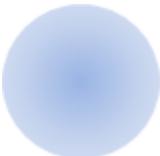 Noun | A word that refers to a person, place, object, event, substance, idea, feeling, or quality. For example, the words "teacher", "book", "development", and "beauty" are nouns. |
| 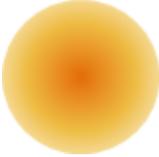 Adjective | a word that describes a noun or pronoun. The words "big", "boring", "purple", and "obvious" are all adjectives. | 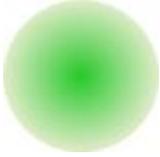 Verb | A word that is used to say that someone does something or that something happens. For example, the words "arrive", "make", "be", and "feel" are verbs. |
| 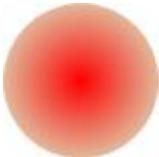 Adverb | A word that describes or gives more information about a verb, adjective, phrase, or other adverb. In the sentences "He ate quickly." and "It was | 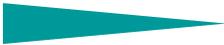 Is a part of | **hyponym/hypernym pair** With regards to "wheat" and "grain", we can understand this to mean that wheat "is a kind of" grain. Here, "wheat" is a hyponym and "grain" is a |
| | extremely good.", "quickly" and "extremely" are both adverbs. | | hypernym. In the case of verbs this can be understood by "is one way to". For example, to trot "is one way to" walk. |
| 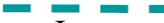 Is an instance of | **hyponym/hypernym pair** In these relationships, the hyponym is specific and unique. For example, "Einstein" is an instance of a "physicist". | 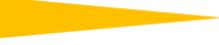 Is a member of 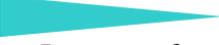 Is a part of 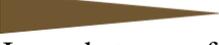 Is a substance of | **meronym/holonym pair** In these cases, the meronym in some way belongs to the holonym. Examples: "robin" is a member of the "thrushes", a "wheel" is a part of a "wheeled vehicle", "caffeine" is a substance of "coffee". |
| 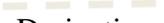 Derivation | These relationships represent the process of forming a new word from an existing word, often by adding a prefix or suffix, such as un- or -ness. For example, "unhappy" and "happiness" derive from the root word "happy". | 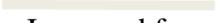 Is a word for | These relationships denote inclusion within a word or phrase. For example, "wheat is a word for wheat germ" |

Instructions of the game are the following:

1. Each player receives a template of a term to complete and N terms to fill it out.
2. The term must be consistent with the definition stated in the template.
3. At the end of the placement of the terms the player goes on to validate his answers. The system displays the score of the related terms appropriately and assigns a star rating.
4. The player ascends to the next level that includes a higher level of difficulty.
5. The highest score determines the winning player.

The general structure of each level is depicted are the following:

- Level 1: contains 4 different synsets networks. Each network is formed by a unique kind of term (noun, verb, adjective or adverb) and a unique kind of association (Is a word for).
- Level 2: contains 2 different synsets networks. Each network is formed by the combination of two kind of terms and two kinds of association.
- Level 3: Contains a unique synsets network. The network is formed by the combination of the 4 kinds of terms and various kinds of association.
- Level 4: Contains a unique synsets network. The network is formed by combinations of terms and relational connections such as level 3 but with greater difficulty.

**5. Conclusions and future work**

Onto-ling is a useful strategy for helping in the teaching-learning process. Students are introduced in linguistics ontologies by learning the basic linguistic elements of any domain. The game is easy and funny, and it allows students for learning about kinds of terms and basics associations. In addition, the game highlights the importance of the definitions and synonyms in linguistics ontologies.

Future work is aimed to improve the game by adding new levels with different compositions. In addition, new definitions and explorations related to new terms are considered. Finally, a detailed specification of created expression according to the kind of association can be applied as an extension of the base game.